\def\bphi{\vec{\phi}}

\def\bpi{\vec{\pi}}

   \documentstyle[12pt]{article}
   
\begin{document}
   \begin{center}
       {\Large\bf The Rotationally Improved Skyrmion, or ``RISKY''}
   \vskip 1.0cm
       {\bf NICHOLAS DOREY}\\
       {\it Physics Department, University of Wales Swansea, } \\
       {\it Singleton Park, Swansea, SA2 8PP, UK}
   \vskip .4cm
       and
   \vskip .4cm
       {\bf  MICHAEL P. MATTIS} \\
       {\it Theoretical Division, Los Alamos National Laboratory,} \\
        {\it Los Alamos, NM 87545, USA}
   \end{center}
   \vskip 1.5cm
   \begin{abstract}
The perceived inability of the Skyrme model to reproduce pseudovector
pion-baryon coupling has come to be known as the ``Yukawa problem.''
In this talk,\footnote{Talk presented at the 1995 International Workshop on
Nuclear and Particle Physics in Seoul, Korea}
 we review the complete solution to this problem.
The solution involves a new configuration known as the rotationally improved
Skyrmion, or ``RISKY,'' in which the hedgehog structure is modified
by a small quadrupole distortion.
We illustrate our ideas both in the Skyrme
model and in a simpler model with a global $U(1)$ symmetry.

   \end{abstract}

\bf Introduction. \rm
The Skyrme model \cite{skyrme,anw} provides an approximate
description of the baryon spectrum of QCD
in the large-$N_{c}$ limit. The semiclassical expansion about the Skyrmion
corresponds directly to the $1/N_{c}$ expansion in the
underlying gauge theory. In the previous talk \cite{preceding},
 we established a
direct connection between effective Lagrangian models, in which baryons
are represented by pointlike Dirac spinor fields, and Skyrme-type
models, in which baryons are instead pictured as
solitons in the field of mesons. The connection is by means of the
large-$N_c$ Renormalization Group; we reviewed under what
circumstances Skyrme-type models are, or are not, the ultraviolet fixed points
of the effective field theories, under the action of this RG flow.

Here we discuss
the equivalence in the \it opposite \rm direction: starting with the Skyrme
model, we explain how it bootstraps itself into an equivalent effective
meson-baryon Lagrangian, with explicit (rather than solitonic)
 baryon fields. In other
words, this talk supplies the ``missing third leg'' of Fig.~1 from
the preceding lecture.

Obviously the  very first effective vertex one needs to recreate is the
pseudovector Yukawa coupling of the pion to the $I=J$ tower of
large-$N_c$ baryons. Such a vertex is needed to account, not only
for the hadronic decay of the $\Delta,$ that is $\Delta\rightarrow
\pi+N,$ but also for the virtual processes $N\rightarrow \pi+N$ or
$\Delta\rightarrow\pi+\Delta.$ Yet reproducing this vertex directly
from Skyrmion quantization has proved to be the most longstanding
headache in the Skyrme-model literature, and has come to be known as
the ``Yukawa problem.''\footnote{Surprisingly, the next-most-complicated
effective vertex, in which two pions are attached to the baryon,
is less sensitive to the Yukawa problem, and indeed constitutes one
of the early phenomenological successes of the Skyrme model
\cite{peskin,HEHW}.} The origin of this problem lies in the fact
that first variations off the static hedgehog Skyrmion vanish, as the
Skyrmion is, by definition, a solution to the Euler-Lagrange equations.
In this talk we review the complete solution to the Yukawa problem,
following our recent work \cite{dhm1,dhm2}.\footnote{For a comprehensive
list of references to
 other approaches to the Yukawa problem, see Ref.~\cite{dhm2}.}
At the same time, we present the solution to another longstanding
problem, not just with the Skyrme model, but with the entire
large-$N_c$ program: the fact that, as $N_c\rightarrow\infty,$ the
spectrum of ground-state baryons contains not only the nucleon and
the $\Delta$ as desired, but also unwanted, unseen states with
$I=J={5\over2},{7\over2},{9\over2},\cdots.$

While tied to the previous talk through Fig.~1, the physics issues involved
here are actually quite different. We will be interested in
the \it analytic properties \rm of Skyrmions, and will find that
while the usual hedgehog Skyrmion is inadequate,
a close relative, which we call the \it rotationally improved Skyrmion\rm,
or RISKY, serves our purpose well. In a nutshell, the RISKY is obtained by
including the (iso)rotational kinetic energy term in the
minimization of the static Hamiltonian.  It is characterized by
 an interesting small
quadrupole distortion away from the hedgehog ansatz (as has been
derived independently, using other physics entirely, by Schroers
\cite{schroers}).

\bf The Yukawa problem. \rm
It has long been suspected that the effective S-matrix element describing
$\Delta\rightarrow\pi+N,$  $N\rightarrow \pi+N$, etc., in the
Skyrme model
 may be expressed in  a particularly compact form. In \cite{anw} it is argued
from general considerations of symmetry and mass dimensions that the
{\it effective} coupling of soft pions to the Skyrmion should be of the
form
\begin{equation}
{\cal L}= \frac{3g_{\pi NN}}{4\pi f_{\pi}M_{N}} \partial_{i}\pi^{a}
{\rm Tr}\,\tau_{i}A^{\dagger}\tau_{a}A\ ,
\label{coupling}
\end{equation}
where $A\in SU(2)$ is the (iso)rotational collective coordinate of the
Skyrmion and the overall normalization of the coupling is determined
by evaluating its matrix element between initial and final nucleon
states. We stress the word {\em effective} because, obviously,
this coupling is not present in the Skyrme Lagrangian itself.
The interaction (\ref{coupling}) yields
predictions for the couplings of pions to the whole $I=J$ tower of
baryons which arise when the isorotational motion of the Skyrmion is
quantized. In particular, the effective vertex for $\Delta$ decay comes with a
coupling constant $g_{\pi N\Delta}=\frac{3}{2}g_{\pi NN}$; this
relation yields a value for the width of the $\Delta$ within a few MeV
of its experimentally measured value. More generally, (\ref{coupling}) embodies
the ``proportionality rule'' which fixes the ratios of the pion
couplings to all the baryons
in the large-$N_c$ $I=J$ tower, and which may be derived in many different ways
 (see preceding lecture for a discussion).

For many reasons, therefore, we should expect that the leading-order
semiclassical S-matrix elements for pion-baryon interactions in the
Skyrme model should coincide with the tree-level  effective
Yukawa couplings implied by (\ref{coupling}). But as mentioned
earlier, the problem of
showing, from first principles of soliton quantization, that this is
actually the case presents some difficulty.
 Before presenting our solution to this Yukawa problem it is
convenient to review some of the basic issues in the context
of $\Delta$ decay. The obvious starting point for calculating the
S-matrix element for $\Delta$ decay is the one-point Green's function
of the pion field evaluated between an incoming $\Delta$-state and an
outgoing nucleon,
$G^{a}({\bf x},t)=\langle N|\pi^{a}({\bf x},t)|\Delta\rangle$.
When quantizing the Skyrmion it is customary to
split up the pion field as $\vec\pi=\vec\pi_{\rm cl}+\delta\vec\pi$ where
$\pi_{\rm cl}^{a}({\bf x},t)=\sin F(r) D_{ab}^{(1)}(A(t))\hat{x}^{b}$ is
the background Skyrmion field configuration and $\delta\vec\pi$ is the
fluctuating part of the pion field. This division of the field defines two
contributions to the one-point function which we now examine in
turn.

 Early approaches to this problem identified the physical pion
field with the fluctuating part $\delta\vec\pi$
\cite{ver}. A contribution of this
sort requires a term linear in $\delta\vec\pi$ in the expansion of the
Skyrme Lagrangian. However, precisely because the static Skyrmion is a
solution of the field equation, such a term is absent at leading order
in $1/N_c$
and only appears when the rotation of the Skyrmion is included. The
resulting coupling is proportional to $\dot{A}$ and
(since the large-$N_c$ Skyrmion (iso)rotates very slowly)
is therefore down in the
 $1/N_{c}$ expansion relative to the desired result (\ref{coupling}).

Because of these difficulties, several authors \cite{dp,japanese}
suggested that the
effective Yukawa interaction comes instead from the classical
contribution to the pion one-point function, $G^{a}_{\rm cl}({\bf x},t)=\langle
N|\pi^{a}_{\rm cl}({\bf x},t)|\Delta\rangle$. However, as it stands, this
idea cannot be correct either. To see why not,
 we must consider the corresponding
momentum space one-point function. Because the Skyrmion profile
function decays exponentially at large $r$, its Fourier transform will
have the following pole contribution:
\begin{equation}
\tilde{G}_{\rm cl}({\bf k},\omega) \sim
\frac{\delta(\omega-M_{\Delta}+M_{N})}
{|{\bf k}|^{2}+m_{\pi}^{2}}\ .
\label{poleI}
\end{equation}
The position of the pole is always at an imaginary value of the pion momentum
$|{\bf k}|=\pm im_{\pi}$. However, so long as
$m_{\pi}<M_{\Delta}-M_{N}$, $\Delta$ decay occurs at the {\em real} value of
the momentum dictated by energy and momentum conservation. Thus the
naive semiclassical Green's function (\ref{poleI}) cannot contribute
to the on-shell S-matrix element via the LSZ reduction formula.  Put another
way, LSZ requires that the denominator in (\ref{poleI})  read
$|{\bf k}|^{2}+m_{\pi}^{2}-\omega_\pi^2$ rather than just
$|{\bf k}|^{2}+m_{\pi}^{2}$, where $\omega_\pi$ denotes the energy
of the emitted pion, or equivalently, the difference between the initial
and final Skyrmion energies.

In summary, neither of the two contributions to the pion Green's
function identified above can, by itself, reproduce the effective
Yukawa vertex (\ref{coupling}). In the remainder of this talk,
we will explain how, when taken together, the Yukawa problem is
solved, in a rather surprising way.

\bf Skyrmion quantization in 3+1 and 1+1 dimensions. \rm
In order to motivate our solution it will be
convenient to consider a simpler model than the $SU(2)$ Skyrme
model which has only an abelian global internal symmetry.  In the following we
will develop both models in parallel for maximum clarity.
We will consider the case of a real two-component scalar field
$\bphi=(\phi_1,\phi_2)$ in two space-time dimensions,
\begin{equation}
{\cal L}=\frac{1}{2}\partial_{\mu} \bphi\cdot
\partial^{\mu} \bphi-\frac{m^{2}}{2}|\bphi|^{2}- W(\bphi)\ .
\label{lag1}
\end{equation}
In contrast, the Skyrme model is described by the following Lagrangian
for an $SU(2)$ valued matrix field in four-dimensional spacetime:
\begin{equation}
{\cal L}=\frac{f^{2}_{\pi}}{16}{\rm Tr}\,\partial_{\mu}U^{\dagger}
\partial^{\mu}U+\frac{1}{32e^{2}}{\rm Tr}\,[U^{\dagger}
\partial_{\mu}U\, , \, U^{\dagger}\partial_{\nu}U]^{2}
+\frac{m_{\pi}^{2}f_{\pi}^{2}}{8}{\rm Tr}\left(U-1\right)\ .
\label{lag2}
\end{equation}
We will choose to rewrite the model in terms of the pion field as
$U=\exp(2i\bpi\cdot \vec{\tau}/f_{\pi})$. The Lagrangian then
takes the general form
\begin{equation}
{\cal L}=\frac{1}{2}\dot{\pi}^{i}g_{ij}(\bpi)\dot{\pi}^{j}-
V(\bpi,\partial_{i}\bpi)
\label{lag3}
\end{equation}
In fact, it will not be necessary to specify the target space metric
$g_{ij}(\bpi)$ or the potential $V(\bpi,\partial_{i}\bpi)$.
Our analysis will apply to any chiral soliton model which can be
written in the above form.

Provided the potential $W$ has its minima at $\bphi=0$,
the model (\ref{lag1}) has an unbroken $U(1)$ symmetry,
$\bphi\rightarrow{\cal M}(\theta)\cdot\bphi$, where
\begin{equation}
{\cal M}(\theta)=
\left(\begin{array}{cc} \cos\theta & \sin\theta \\ -\sin\theta &
\cos\theta \end{array} \right)
\label{rot}
\end{equation}
Correspondingly, the Skyrme model has an unbroken $SU(2)$ symmetry
$U\rightarrow AUA^{\dagger}$. In terms of the pion field of
(\ref{lag3}) this symmetry takes the form $\bpi\rightarrow
D^{(1)}(A)\cdot\bpi$ where
\begin{equation}
D^{(1)}_{ij}(A)=\frac{1}{2}{\rm
Tr}\,\tau_{i}A\tau_{j}A^{\dagger}\ .
\end{equation}

In either theory, soliton solutions are found by minimizing the
corresponding mass functionals:
\begin{eqnarray}
\mu[\bphi] & = & \int dx\, \frac{1}{2}|\bphi'|^{2}+
\frac{m^{2}}{2}|\bphi^{2}|+W(|\bphi|) \ ,\nonumber \\
M[\bpi] & = & \int d^{3}{\bf x}\, V(\bpi,\partial_{i}\bpi) \ .
\end{eqnarray}
In the $U(1)$ model we will assume that this yields a soliton solution
of the form $\bphi^{\rm cl}=(\phi_{S}(x),0)$ \footnote{Actually, in order
to get such a soliton it is necessary to include in the model an
additional scalar field \cite{raj}. However, this field is neutral
under the $U(1)$ symmetry and its presence does not affect our analysis
in any way}. For reasons that will become
clear below, we will sometimes write the profile function as $\phi_{S}(x;m)$ to
emphasize its parametric dependence on the meson mass $m$.
In the Skyrme model the
Euler-Lagrange equation $\delta M[\bpi]/\delta \pi^{a}=0$ yields a
chiral soliton solution with the characteristic hedgehog form
\begin{equation}
{\bpi}_{\rm cl}=\frac{f_{\pi}}{2}\sin F(r)\, \hat{\bf r}\ .
\end{equation}

In both models the ansatz chosen is the one of maximal symmetry; our
assumption that the $U(1)$ soliton can be chosen to lie entirely in
the first component of the field $\bphi$ is analogous to the hedgehog
ansatz in this respect. The profile functions,
$\phi_{S}(x)$ and $F(r)$ must be determined by solving a non-linear
ODE. However, in both cases, the spatial asymptotics of these
functions can be determined by solving the corresponding linearized
equation. Thus $\phi_{S}(x;m)\rightarrow A\exp\left(-mx\right)$ as
$|x|\rightarrow\infty$, while the large-$r$ asymptotics of $F$ are
given by
\begin{equation}
F(r)\rightarrow B\cdot\left(\frac{m_{\pi}}{r}+\frac{1}{r^{2}}\right)
\exp\left(-m_{\pi}r\right)\ .
\label{asymptotics}
\end{equation}
The constants $A$ and $B$ must determined by solving the
corresponding non-linear equation numerically.

In the $U(1)$ model the full set of static one-soliton solutions is
obtained by acting on the solution $\bphi^{\rm cl}$ with a $U(1)$ rotation
and a translation:
\begin{equation}
\bphi^{\rm cl}(x;X,\theta)={\cal M}(\theta)\cdot\bphi^{\rm cl}(x-X)
\label{fullset1}
\end{equation}
while, for the Skyrme model, the full set is parametrized by the
collective coordinates $A\in SU(2)$ and ${\bf X}\in {\cal R}^{3}$:
\begin{equation}
\bpi^{\rm cl}\left({\bf x};{\bf X},A\right)=D^{(1)}(A)\cdot\bpi^{\rm cl}
\left({\bf x}-{\bf X}\right)
\label{fullset2}
\end{equation}
We are ignoring the Lorentz contraction of the soliton for
notational simplicity; moreover
the translational collective coordinates
will not play an important role in what follows
and we will suppress them below. Instead
 we will concentrate on the dynamics of the internal
collective coordinates $\theta$ and $A$.

It is convenient to define,
in each model, a moment of inertia which is a functional of the field:
\begin{eqnarray}
\lambda[\bphi] & = & \int dx\, \bphi\cdot\bphi  \nonumber \\
\Lambda_{ij}[\bpi] & = &\int d^{3}{\bf x}\,
\epsilon_{abi}\pi^{a}g_{bd}(\bpi)\epsilon_{jcd}\pi^{c}
\label{MI}
\end{eqnarray}
In both models the semi-classical baryon spectrum is obtained by
allowing the rotational collective coordinate to be time-dependent. In
the Skyrme case the effective Lagrangian for this degree of freedom
is\footnote{When evaluated on a hedgehog, the moment of inertia tensor
collapses to a scalar: $\Lambda_{ij}=\Lambda\,\delta_{ij}.$ Deviations
from hedgehog structure only affect this result at higher order
in $1/N_c.$}
\begin{equation}
L=-M+\Lambda{\rm Tr}\,\dot{A}\dot{A}^{\dagger}
\label{anwlag}
\end{equation}
where both $M$ and  $\Lambda\sim N_{c}$.
 Quantizing this system gives the standard rotor spectrum
first obtained by Adkins, Nappi and Witten: an infinite tower of
states $|I=J,i_{z},s_{z}\rangle$ with masses,
\begin{equation}
M(J)=M+\frac{J(J+1)}{2\Lambda}
\label{anwspectrum}
\end{equation}
In the $U(1)$ model it is trivial to carry out the same procedure; in
this case the effective dynamics of the time-dependent angle
$\theta(t)$ is just that of
a free particle moving on a circle,
\begin{equation}
L=-\mu+\frac{1}{2}\lambda\dot{\theta}^{2}\ .
\end{equation}
In order to exploit the analogy to the Skyrme model to
the full, we will choose the $N_{c}$ dependence of
the coupling constants in (\ref{lag1}) so that $\mu\sim N_{c}$ and
$\lambda\sim N_{c}$ (recognizing that in this toy model $N_c$ no longer
corresponds to ``quark number'' of any sort, but merely parametrizes
the semiclassical expansion).
The resulting spectrum of states is labeled by a
conserved $U(1)$ charge $q=0,\pm 1,\pm 2\ldots$ analogous to the
(iso)spin quantum numbers of the Skyrme model baryon states. The
corresponding mass spectrum and wave-functions are given by,
\begin{eqnarray}
\mu(q)=\mu+\frac{q^{2}}{2\lambda} & \qquad{} \qquad{} &
\langle q|\theta\rangle=\frac{1}{\sqrt{2\pi}}\exp(iq\theta)
\label{u1spectrum}
\end{eqnarray}

\bf Rotationally improved Skyrmions. \rm
In the $U(1)$ model there is a process which is precisely analogous to
$\Delta$ decay: the state $|q+1\rangle $ can decay on shell to the state
$|q\rangle$ with the emission of a single physical meson, provided, of course,
that the meson mass $m$ is less than the splitting between these
states. Before describing how the S-matrix element for $\Delta$ decay
can be calculated in the Skyrme model, we will consider this simple
process which contains many of the features of the Skyrme
case. As described earlier, the relevant Green's function,
from which the S-matrix element can be extracted, is the one point
function of the meson field sandwiched between the initial and final
soliton states.
\begin{equation}
g_{a}(x,t)=\langle q|\phi_{a}(x,t)|q+1\rangle
\label{green}
\end{equation}
The key idea introduced in Ref.~\cite{dhm1} is that this Greens
function is given to leading order in the semiclassical approximation
by replacing $\bphi$ with a saddle-point field configuration
$\bphi_{\rm sp}(x;\theta,q)$ which depends both on the collective
coordinate $\theta$ of the soliton and the conserved $U(1)$ charge
$q$. The relevant saddle-point equation is obtained by minimizing the
the mass functional $\mu[\bphi]$ augmented by a correction term due to
the rotational motion of the soliton:
\begin{equation}
\frac{\delta}{\delta\phi_{a}} \left( \mu[\bphi] +
\frac{q^{2}}{2\lambda[\bphi]}\right)=0
\end{equation}
This works out to,
\def\sqr#1#2{{\vcenter{\vbox{\hrule height.#2pt
        \hbox{\vrule width.#2pt height#1pt \kern#1pt
           \vrule width.#2pt}
        \hrule height.#2pt}}}}
\def\square{\mathchoice\sqr84\sqr84\sqr53\sqr43}
\begin{equation}
(\square+m^{2})\phi_{a}+\frac{\delta W}{\delta
\phi_{a}}-\frac{q^{2}\phi_{a}}{\lambda^{2}[\bphi]}=0
\label{saddle}
\end{equation}
The only modification, therefore, of the static equation of motion is
a shift in the meson mass term; $m^{2}\rightarrow
m^{2}-q^{2}/\lambda^{2}[\bphi]$. It is straightforward to relate the
solution of (\ref{saddle}) to the solution $\phi_{S}(x;m)$ of the
static field equation. Writing
$\bphi_{\rm sp}={\cal M}(\theta)\cdot(\varphi,0)$, the saddle-point
profile function is given by
$\varphi(x)=\phi_{S}(x,\sqrt{m^{2}-q^{2}/\lambda^{2}[q]})$ where
$\lambda[q]$ must be determined self-consistently:
\begin{equation}
\lambda[q]=\int dx\, \phi_{S}^{2}(x;\sqrt{m^{2}-q^{2}/\lambda^{2}[q]})
\label{consistent}
\end{equation}

In \cite{dhm1} we showed that, with certain very mild assumptions,
this equation always has a real solution for $q$ sufficiently small.
For the present purposes it is only necessary to observe that the
self-consistency equation can be expanded in powers of $1/N_{c}$
giving $\lambda[q]=\lambda\cdot[1+O(1/N_{c})]$. The net result of this
shift in the effective meson mass is that the profile of the
rotationally improved soliton has a slower exponential fall off at
large $|x|$ than its static counterpart,
\begin{equation}
\varphi(x)\rightarrow A\exp\left(-\sqrt{m^{2}-q^{2}/\lambda^{2}}\,
|x|\right)
\label{slower}
\end{equation}
This ``swelling'' of the soliton is just its response to the centrifugal
forces produced by its internal rotation.

Just as in the simple $U(1)$ case described above, the leading-order
Green's function which contributes to $\Delta$ decay is dominated by a
single field configuration; the Rotationally Improved Skyrmion or
RISKY which satisfies a modified equation of motion,
\begin{equation}
\frac{\delta}{\delta\pi^{a}} \left(M[\bpi]+
\frac{1}{2}J^{m}\Lambda^{-1}_{mn}[\bpi]J^{n}\right)=0
\label{riskeq}
\end{equation}
where ${\bf J}$ is the classical angular momentum vector of the
Skyrmion. Although we cannot solve this equation analytically it is
straightforward to find the leading modification of the spatial
asymptotics of the static Skyrmion. Writing
$\bpi_{\rm sp}=D^{(1)}(A)\cdot{\bpi}({\bf x};{\bf J})$, as
$r\rightarrow\infty$ we have,
\begin{eqnarray}
\bpi({\bf x},{\bf J}) & \rightarrow & \frac{B}{J^{2}} \cdot  \left(
\frac{m_{\pi}}{r}+\frac{1}{r^{2}}\right)\exp\left(-m_{\pi}r\right)\left({\bf
J}\cdot\hat{\bf r}\right){\bf J}  \nonumber \\
&  & \ \  + \frac{B}{J^{2}}\cdot
\left(\frac{\sqrt{m^{2}_{\pi}-J^{2}/\Lambda^2}}{r}+\frac{1}{r^{2}}\right)
\exp\left(-\sqrt{m^{2}_{\pi}-J^{2}/\Lambda^2}\cdot r\right)\left({\bf
J}\times \hat{\bf r}\times{\bf J}\right) \nonumber \\
\label{mf}
\end{eqnarray}
which properly collapses to the hedgehog in the limit $J^2\rightarrow0$,
as the reader may verify.
Here the situation is somewhat more complicated than the $U(1)$ case;
the RISKY field configuration is a superposition of two different
tensor structures. The coefficient of the first tensor structure in
(\ref{mf}), which is parallel to ${\bf J}$, has the same exponential
fall-off as the static Skyrmion. In contrast the coefficient of
the second tensor structure, which is perpendicular to ${\bf J}$ has a
modified tail analogous to that of the rotating $U(1)$ soliton: the
pion mass is shifted as $m_{\pi}^{2}\rightarrow
m_{\pi}^{2}-J^{2}/\Lambda^{2}$. The physical interpretation is clear;
the rotating Skyrmion ``swells'' in the directions perpendicular to
its axis of rotation due to centrifugal forces.

We are still only half way to evaluating the required Green's
function; we have calculated the saddle-point field configuration
which gives the dominant semiclassical contribution for a Skyrmion
with collective coordinate $A$ and conjugate
angular momentum ${\bf J}$, but it is still necessary to quantize
these collective coordinate degrees of freedom. Again we begin by
treating the simpler $U(1)$ case. The collective coordinate $\theta$
and its conjugate momentum $q$ become quantum operators
$\hat{\theta}$, $\hat{q}$ with $[\hat{\theta},\hat{q}]=i$. The problem
of evaluating the leading semiclassical contribution to the
Green's function (\ref{green}) reduces to that of
calculating a quantum mechanical expectation value of the saddle point
field; $g^{i}(x,t)=\langle q | \phi^{i}_{\rm sp}(x,t;\hat{\theta},\hat{q})
|q+1\rangle +O(1/N_{c})$ which gives
\begin{equation}
g^{i}(x,t)=\langle q|{\cal M}_{i1}(\hat{\theta})\cdot
\phi_{S}(x;\sqrt{m^{2}-\hat{q}^{2}/\lambda^{2}})|q+1\rangle
\label{spfo}
\end{equation}
However, as it stands, this expression is ambiguous; we need to
specify an ordering prescription for the non-commuting operators
$\hat{\theta}$ and $\hat{q}$. This ordering
problem is quite generic to soliton quantization where the
introduction of collective coordinates always involves a nonlinear
change of variables involving both the coordinates and their conjugate
momenta \cite{gerv}.
The standard resolution of this problem is to choose the Weyl
ordering prescription which is known to have certain desirable
properties; in the case of translational motion of a soliton in one
dimension this prescription is required to preserve Lorentz
invariance \cite{tomb}. Using standard identities,
the net result of Weyl ordering
the saddle-point field operator in (\ref{spfo}) is that $\hat{q}$ can
be replaced everywhere by its midpoint value $\bar{q}=(q+1/2)$.

Taking a Fourier transform,
the resulting semiclassical Green's function for the
emission of a positively charged meson with energy $\omega$ and
momentum $k$ contains a pole contribution which is dictated by the
spatial asymptotics of $g^{i}(x,t)$.
\begin{equation}
\tilde{g}(k,\omega)=\delta(\omega-\mu(q+1)+\mu(q)) \cdot \frac{2iAk}
{k^{2}+m^{2}-\bar{q}^{2}/\lambda^{2}} \,  +\, \hbox{Non-pole terms}
\label{pole}
\end{equation}
In order for a non-vanishing contribution to the S-matrix it is necessary that
the pole position coincides with the meson mass shell condition
$k^{2}+m^{2}=\omega^{2}=(\mu(q+1)-\mu(q))^{2}$. This follows immediately
because
$\mu(q+1)-\mu(q)=((q+1)^{2}-q^{2})/2\lambda=\bar{q}/{\lambda}$. Hence we
see that, when the operator ordering problem is correctly resolved,
the rotational improvement of the soliton profile has the effect of
shifting the meson pole to exactly the position required by the LSZ
reduction formula. This means that there will be a leading order
contribution to the S-matrix for the decay process $|q+1\rangle
\rightarrow |q \rangle\, +$ one meson which coincides exactly with
the expected Yukawa vertex.

In the Skyrme case, the one-pion Green's function is dominated by the
RISKY, which solves equation
(\ref{riskeq}). By analogy with our treatment of the $U(1)$ case
above, we will now examine the pole contribution to the Fourier
transform of this field configuration which is dictated by the
asymptotic form (\ref{mf}),
\begin{equation}
\tilde{\bpi}({\bf k};\hat{{\bf J}})\sim
\frac{ 4\pi iB}{|{\bf k}|}\left[ \frac{1}{
|{\bf k}|^{2}+m_{\pi}^{2}} \frac{(\hat{\bf J}\cdot {{\bf
k}})\hat{\bf J}}{\hat{J}^{2}}+ \frac{1}{
|{\bf k}|^{2}+m_{\pi}^{2}-\hat{J}^{2}/\Lambda^{2}}
\frac{\hat{\bf J}\times {{\bf
k}}\times \hat{\bf J}}{\hat{J}^{2}}\right]
\label{mf2}
\end{equation}
The resulting momentum-space configuration has
two separate poles, corresponding to the two tensor structures which
contribute to the RISKY. However, this is appropriate in the Skyrme
case because there are two separate processes allowed by spin and isospin
conservation. As well as the real process of $\Delta$ decay,
$\Delta\rightarrow N+\pi$ there are also virtual processes
$N\rightarrow N+\pi$ and $\Delta\rightarrow \Delta+\pi$, etc.
 As we showed in detail in \cite{dhm2} there
exists an ordering prescription for the non-commuting operators
$\hat{\bf J}$ and $\hat{A}$, analogous to the Weyl ordering chosen
in the $U(1)$ case, which produces exactly the required analytic
structure for the saddle-point Green's function evaluated between
states of initial and final spin $J$ and $J'$,
\begin{equation}
\tilde{\bf G}_{\rm sp}({\bf k}, \omega)=4\pi i B {\bf k}
\cdot \left[ \frac{\delta_{J,J'} \delta(\omega)}{|{\bf
k}|^{2}+m_{\pi}^{2}} +\frac{\delta_{J,J'\pm1} \delta(\omega-M(J)+M(J'))}{|{\bf
k}|^{2}+m_{\pi}^{2}-(M(J)-M(J'))^{2}}\right]
\label{end}
\end{equation}
Thus we see that the two tensor structures give poles
corresponding exactly to the two allowed processes.  The linear
dependence on $\bf k$ means pseudovector coupling as desired---even
beyond the soft-pion kinematic regime where this is required by
Adler's rule. Once again, large-$N_c$ reasoning allows an extrapolation
to higher energies.  The constant $B$ is fixed by
(\ref{coupling}) to be $3g_{\pi NN}/8\pi M_N.$

\bf Width calculations and the unwanted $I=J$ baryons. \rm
In sum, we have introduced a new configuration, the RISKY, whose
analytic properties are precisely such that the Skyrme model maps
onto an effective meson-baryon Lagrangian (at least at the level
of the Yukawa vertex (\ref{coupling})). Furthermore, armed with this effective
coupling, we can also confront an important phenomenological objection
to large-$N_c$ physics, namely the existence of the unwanted
$I=J$ baryons with $I\ge{5\over2}.$ They are simply too broad to
be seen! We can address this issue precisely because Eq.~(\ref{coupling})
may be sandwiched between any of the $I=J$ baryons, not just the
nucleon or $\Delta.$
Furthermore, all these decay amplitudes will be proportional to
$g_{\pi NN}$ which sits out in front (the ``proportionality rule''
described in the previous lecture). We refer the interested reader
to Ref.~\cite{dhm2} for the (non-illuminating)
calculational details, and conclude by summarizing the principal findings:

1. With $g_{\pi NN}$ drawn from experiment, the width of the $\Delta$
works out to 114$\,$MeV in the Skyrme model,
 within a few MeV of the actual value.

2. With this same value for $g_{\pi NN}$, the widths of the higher-spin
baryons rises rapidly, thus $\Gamma_{5/2}\,\sim\,800\,$MeV,
$\Gamma_{7/2}\,\sim\,2600\,$MeV,
$\Gamma_{9/2}\,\sim\,6400\,$MeV, etc.
These are so broad that
\it there is no conflict whatever with phenomenology. \rm


\begin{thebibliography}{99}
\bibitem{skyrme} T. H. R. Skyrme, {\em Proc. Roy. Soc.} {\bf A260} (1961) 127.
\bibitem{anw} G. Adkins, C. Nappi and E. Witten, {\em Nucl. Phys. }
{\bf B228} (1983) 552.
\bibitem{preceding} N. Dorey and M. P. Mattis, \it The large-$N_c$
renormalization group\rm, to appear in the Proceedings of the 1995
International Workshop on Nuclear \& Particle Physics: Chiral
Dynamics in Hadrons \& Nuclei, hep-ph/9504328.
\bibitem{peskin} M. Mattis and M. Karliner {\em Phys. Rev.} {\bf D31} (1985)
2833;
 M. Mattis and M. Peskin, {\em Phys. Rev.} {\bf D32} (1985) 58.
\bibitem{HEHW} A. Hayashi, G. Eckart, G. Holzwarth and H. Walliser,
\it Phys.~Lett.~\bf B147 \rm (1984) 5.
\bibitem{dhm1}  N. Dorey, J. Hughes and M. P. Mattis, {\em Phys. Rev.}
{\bf D49} (1994) 3598.
\bibitem{dhm2} N. Dorey, J. Hughes and M. P. Mattis,
{\em Phys. Rev.} {\bf D50} (1994) 5816.
\bibitem{schroers} B. J. Schroers, \it Zeit.~Phys.~\bf C61 \rm (1994) 479.
\bibitem{ver} H. Verschelde, {\em Phys. Lett.} {\bf B209} (1988) 34;
 G. Holzwarth, A. Hayashi and B. Schwesinger, {\em Phys. Lett.}
{\bf B191} (1987) 27;
 S. Saito, {\em Prog. Theor. Phys.} {\bf 78} (1987) 746.
\bibitem{dp} D. I. Dyakonov, V. Yu. Petrov, and P. B. Pobylitsa,
{\em Phys. Lett.} {\bf B205} (1988) 372.
\bibitem{japanese} A. Hayashi, S. Saito and M. Uehara,
{\em Phys. Lett.} {\bf B246} (1990) 15;
 \sl Phys. Rev. \bf D43\rm, 1520 (1991); \it ibid\rm., \bf D46\rm, 4856
(1992); \em Prog.~Theor.~Phys.~Supp.~\bf109 \rm (1992) 45.
\bibitem{raj} R. Rajaraman and E. Weinberg, {\em Phys. Rev.} {\bf D11}
(1975) 2950.
\bibitem{gerv} J. Gervais and  A. Jevicki {\em Nucl. Phys.} {\bf B110}
(1976) 93.
\bibitem{tomb} E. Tomboulis, {\em Phys. Rev.} {\bf D12} (1975) 1678.


\end{thebibliography}
\end{document}